# Metrics, KPIs, and Taxonomy for Data Valuation and Monetisation - Internal Processes Perspective

EDUARDO VYHMEISTER, University College Cork, Ireland
BASTIEN PIETROPAOLI, University College Cork, Ireland
ALEJANDO MARTINEZ MOLINA, Centro Tecnológico de Investigación, Desarrollo e Innovación en tecnologías de la Información y las Comunicaciones - ITI, Spain
MONTSERRAT GONZÁLEZ-FERREIRO, EGI Fundation, The Netherlands
GABRIEL GONZÁLEZ-CASTANE, Big Data Value Association, Belgium
JORDI ARJONA AROCA, Centro Tecnológico de Investigación, Desarrollo e Innovación en tecnologías de la Información y las Comunicaciones - ITI, Spain
ANDREA VISENTIN, University College Cork, Ireland

Data valuation and monetisation are emerging as central challenges in data-driven economies, yet no unified framework exists to measure or manage data value across organisational contexts. This paper presents a systematic literature review of metrics and key performance indicators (KPIs) relevant to data valuation and monetisation, focusing on the Internal Processes Perspective of the Balanced Scorecard (BSC). As part of a broader effort to explore all four BSC perspectives, we identify, categorise, and interrelate hundreds of metrics within a comprehensive taxonomy structured around three core clusters: Data Quality, Governance & Compliance, and Operational Efficiency. The taxonomy consolidates overlapping definitions, clarifies conceptual dependencies, and links technical, organisational, and regulatory indicators that underpin data value creation. By integrating these dimensions, it provides a foundation for the development of standardised and evidence-based valuation frameworks. Beyond its theoretical contribution, the taxonomy supports ongoing practical applications in decision-support systems and data valuation models, advancing the broader goal of establishing a coherent, dynamic approach to assessing and monetising data across industries.

---

# 1 INTRODUCTION

With the ongoing digitisation of information and the broader digital transformation of organisations, data has emerged as a cornerstone asset across virtually all industries [109, 122, 135]. Firms are increasingly allocating resources to data collection, storage, and analytics, recognising that high- quality data underpins innovation and constitutes a major source of competitive advantage.

As organisations continue to treat data as a key strategic resource, the cumulative effect of these efforts is evident at the microeconomics level, but also across the wider economy. The strategic investment in data infrastructure, advanced analytics, and digital capabilities have contributed to a rapidly expanding global data ecosystem, where data-driven innovation now plays a central role in shaping industry dynamics and economic growth. This macro-level transformation is reflected in the accelerating scale of the data economy, with countries and regions increasingly recognising data as a fundamental pillar of competitiveness and digital sovereignty.

The global data economy is now valued in the hundreds of billions of euros. In 2023 alone, the estimated value was around €350B in the United States, €82B in the European Union, €53B in Japan, and €50B in China, with continued acceleration projected in the coming years [36]. However, while the macro-level growth signals the strategic and economic relevance of data, realising this value in practice, both at the organisational level and across wider markets, remains far from straightforward [99, 135].

Bridging the gap between data availability and measurable financial or operational impact requires deliberate mechanisms, governance models, and managerial capabilities. It is in this context that the concept of data monetisation has gained prominence, not only as an emerging academic domain but also as a practical managerial and strategic concern for firms and policymakers. Although increasingly recognised across research and industry alike, data monetisation still lacks a universally accepted definition and is interpreted in multiple ways, reflecting its evolving nature and diverse applications [77, 90, 103, 109, 135].

To advance conceptual clarity and guide implementation, scholars and practitioners typically classify data monetisation according to the pathways through which value is generated. One of the most common distinctions relates to the strategic orientation of monetisation mechanisms, broadly differentiating between internal and external approaches. Among the widely recognised categories are:

- Internal data monetisation, improving decision-making and operational efficiency to generate financial benefits (e.g., manufacturing cost optimisation);
- Indirect monetisation, where data enhances the value of products or services (e.g., recommendation systems); and
- Direct external monetisation, involving the commercial exchange of data (e.g., through data brokers).

As firms navigate these different routes, a central question emerges: how should data be valued in order to support and justify monetisation decisions? The ability to define, measure, and communicate the value generated, or the value expected from data assets becomes essential for prioritising investments, pricing data-driven products, and demonstrating returns.

The perceived value of data is highly context-dependent and varies with the chosen monetisation approach, making it difficult for organisations to assign a consistent or "fair" value [99]. In the absence of formal valuation frameworks, practices remain fragmented, often producing inconsistent results. Developing robust and standardised methodologies for valuing and monetising data is therefore an urgent priority.

Defining a clear strategy for data valuation and monetisation is essential. The literature highlights several approaches, including Data-as-a-Service (DaaS), Information-as-a-Service (IaaS), servitisation, and the creation of data-driven products [77, 90]. Organisations often adopt multiple strategies simultaneously, depending on their strategic objectives. For example, a firm may leverage internal data to optimise production processes and reduce costs while also deploying customer data to power a recommender system in its online store.

Whether for valuation or monetisation strategies, organisations typically rely on metrics and key performance indicators (KPIs) to assess their performance. Data quality, measured by attributes such as accuracy, completeness, or timeliness, has been widely applied in existing frameworks [97, 98, 128]. However, quality is only one dimension of a broader strategy. Other relevant measures include acquisition and storage costs, accessibility and retrievability in case of failure, or fairness and bias in the data collected. Depending on organisational goals and stakeholder perspectives, these additional metrics may prove equally central to determining both the value and monetisation potential of data.

A wide range of metrics and KPIs have been proposed in the literature to address different dimensions of organisational performance [156]. In this survey, we concentrate on the *internal processes perspective* and examine the metrics and KPIs relevant to both data valuation and data monetisation strategies. Our contributions are two-fold: 1) we provide an in-depth and expansive literature review on metrics and KPIs from the internal processes perspective, 2) we propose a new taxonomy for these metrics and KPIs based on the Balanced Scorecard (BSC) [69].

This paper is organised as follows: Section 2 provides the necessary background to understand this study including the high-level taxonomy we are using; Section 3 reviews metrics and KPIs used in measuring data quality; Section 4 reviews metrics and KPIs that can be used for governance and compliance; Section 5 reviews metrics and KPIs used in measuring operational efficiency; Section 6 discusses our findings and the challenges the domain is facing; and finally, Section 7 concludes the paper.

## 2 BACKGROUND

This section provides the background necessary to situate our work. We review data valuation and monetisation strategies, highlight the role of metrics and KPIs, and describe the methodology of our systematic literature review. Finally, we introduce the Balanced Scorecard (BSC), which we leverage to construct our taxonomy of metrics and KPIs.

### 2.1 Data Valuation and Data Monetisation

*Data valuation* refers to estimating the worth of data, whether expressed in monetary or non-monetary terms. By contrast, *data monetisation*, defined by Gartner as "the process of using data to obtain quantifiable economic benefit" [52], encompasses the practical mechanisms by which data generates value. While valuation is typically conducted internally, monetisation may occur through internal use, indirect enhancement of offerings, or direct external commercialisation.

The notion of *data value* varies depending on the domain or context [11]. Broadly, it refers to the perceived worth of data, whereas *data price* denotes its explicit monetary cost (e.g., the selling price of a dataset). Importantly, data value is inherently subjective: a dataset may hold very different value for distinct stakeholders depending on context, use case, and strategic priorities.

Thus, data value can be regarded as an internal estimate of a dataset's utility or impact within an organisation, while data price represents the external, transactional value set in a marketplace. As with other assets, organisations will consider purchasing data when their internal valuation exceeds the seller's asking price.

## 2.2 Strategies

A business strategy defines the long-term plan through which an organisation seeks to achieve sustainable competitive advantage. It shapes resource allocation, priority-setting, and decision-making in line with the organisation's mission and market positioning. For example, a firm pursuing sustainability may prioritise eco-innovation and green operations, which in turn influence its choice of key performance indicators.

In the context of data, strategy determines how organisations leverage their information assets. The literature identifies three broad categories of data monetisation: internal monetisation, indirect external monetisation, and direct external monetisation [77, 90, 103, 109, 135]. These categories capture whether value is derived through efficiency gains, enhanced services, or direct data sales. Ofulue et al. [103] propose a framework identifying multiple monetisation types, associated services, revenue models, and stakeholders. Similarly, report [77] classifies strategies according to data flow, business orientation (internal vs external), and cash flow. Their analysis identifies 12 strategies that broadly encompass the approaches described in the wider literature [44, 103, 109, 122, 135, 141]. We refer the interested reader to [77] for detailed coverage.

These strategic orientations are not only relevant for categorising monetisation pathways, they also establish the organisational context in which data-related performance must be understood and evaluated. In practice, strategy determines which data assets are prioritised, the capabilities required to govern and exploit them, and the organisational processes through which value is generated and captured. Consequently, strategic choices directly shape the selection of metrics and KPIs by identifying the factors that matter most for performance, whether related to data quality, operational efficiency, customer value, regulatory compliance, or revenue realisation. In other words, strategy provides the contextual lens that translates high-level data ambitions into measurable indicators, ensuring that data and data-management activities are monitored not in isolation but in alignment with the value-creation objectives they are intended to support.

## 2.3 KPIs and Metrics

Metrics and Key Performance Indicators (KPIs) are both used to evaluate organisational performance, but differ in scope and strategic relevance.

*Metrics* represent general-purpose measurements, often used in daily operations to monitor processes, detect trends, and guide improvements. They may be objective (e.g., sensor readings) or subjective (e.g., user-reported assessments), and either qualitative (e.g., data clarity) or quantitative (e.g., dataset downloads) [11].

*KPIs*, in contrast, are high-level indicators directly tied to strategic objectives. Typically derived from metrics, KPIs are constrained by explicit benchmarks or targets. They are fewer in number, strategically selected, and must be regularly reviewed and updated to remain aligned with evolving business goals. Crucially, targets should be time-bound (e.g., quarterly, annually) to ensure accountability. For instance, a KPI such as "achieve X% customer satisfaction by year-end" provides a measurable, time-framed objective that aligns organisational efforts with strategy.

While data quality metrics are essential, it is important to emphasise that the metrics and KPIs used throughout this work are not restricted solely to datasets. Because data monetisation concerns the creation, delivery, and capture of value from data-driven assets and services, performance indicators must extend to processes, systems, governance mechanisms, and business outcomes. Accordingly, metrics in this taxonomy evaluate not only the intrinsic characteristics of data, but also the operational efficiency of data pipelines, contractual and ownership structures, compliance and trust frameworks, and the realised economic value of data-enabled products and services. This distinction reinforces that valuation and monetisation require a holistic lens.

## 2.4 Systematic Literature Review

This survey contributes to a larger systematic literature review conducted using targeted keyword searches [112, 113], structured according to the PICOC methodology [23]. We also employed snowballing, particularly by leveraging existing review articles as entry points (e.g., [11]). Further methodological details are available in [156].

## 2.5 Balanced Scorecard

The Balanced Scorecard (BSC) is a framework designed to align daily operations with long-term strategic goals [69]. Originally developed in the 1990s to complement financial indicators with non-financial measures, it has since become a widely adopted tool for integrating strategy with performance management.

The BSC comprises four perspectives:

- *Financial*: Evaluates how strategic initiatives impact financial outcomes (e.g., revenue growth, profit margins, ROI).
- *Customer*: Focuses on customer satisfaction and loyalty through metrics such as market share and brand recognition.
- *Internal Processes*: Examines operational efficiency and effectiveness, including product development timelines and quality standards. This is the focus of the present paper.
- *Learning & Growth*: Captures organisational capacity for innovation and adaptation, e.g., employee skills, training, and knowledge sharing.

By grouping KPIs under these perspectives, the BSC provides a structured, holistic view of performance. It also offers flexibility for developing data-driven strategies, including those aimed at data monetisation, by ensuring that metrics and KPIs are explicitly linked to broader organisational objectives.

## 2.6 Taxonomy

Drawing on the BSC's four perspectives, we developed a taxonomy [156] to organise and link the metrics and KPIs identified in the literature. Figure 1 presents the high-level structure, comprising the overall business strategy, the four perspectives, and their sub-clusters. The complete taxonomy is available in Appendix A.

The taxonomy is structured across three levels: (i) high-level KPIs for executives and data owners, (ii) mid-level KPIs for departmental strategy implementation, and (iii) low-level metrics for operational specialists. This layered approach facilitates the alignment of strategic goals with actionable KPIs across all levels of the organisation.

In this paper, we concentrate on the *internal processes perspective* and the metrics classified within it. The remaining perspectives will be addressed in future work. For clarity, data centre efficiency metrics are grouped into a dedicated sub-cluster. Although they form an integral part of the broader *operational efficiency* cluster, their extensive scope places a full treatment beyond the focus of this paper. A comprehensive list of these metrics, along with the corresponding references, is provided in Appendix B.

## 3 DATA QUALITY

As organisations seek to monetise data, ensuring its quality becomes a foundational requirement. High-quality data is not merely a technical asset but a strategic enabler: it underpins trust, supports reliable decision-making, and conditions the feasibility of valuation and monetisation mechanisms. Poor-quality data increases operational costs, weakens analytical outcomes, and undermines both internal and external value extraction, limiting the ability to deploy data-driven services, maintain regulatory compliance, or secure customer confidence.

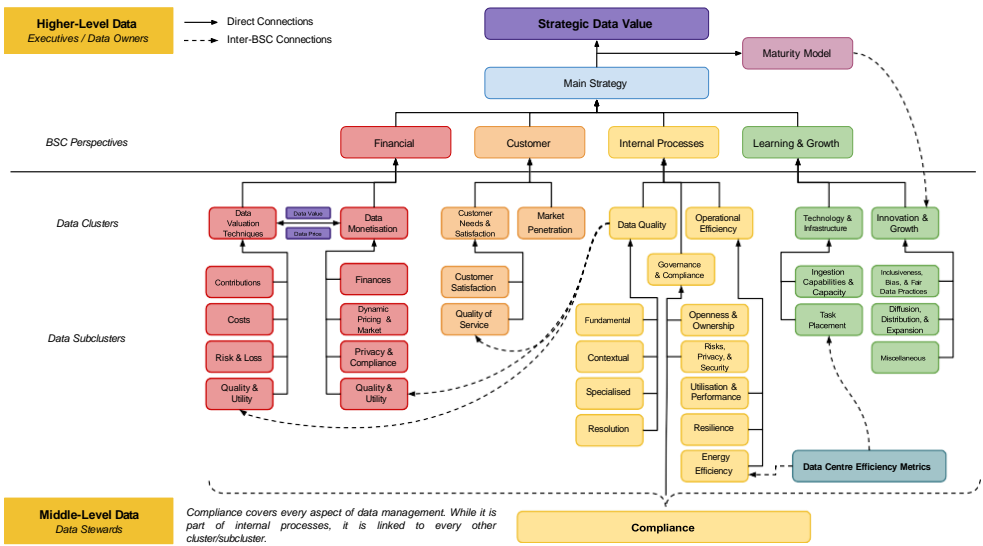

Fig. 1. High-level taxonomy [156]. Each perspective is divided into clusters and sub-clusters used to categorise metrics and KPIs. Dashed lines indicate inter-cluster links. The BSC perspectives and clusters can be aggregated into higher-level KPIs using weighted averages. Data centre metrics, due to their specificity, are categorised in a dedicated sub-cluster. A lower level taxonomy can be seen in Figure 2.

Accordingly, data quality is not an isolated property of datasets it is an organisational capability that shapes market readiness, determines monetisation potential, and ultimately influences the economic and strategic impact of data initiatives. In fact, as observed in the taxonomy (see Figure 1), it can affect the overall valorization approach. This relevance explains why *Data Quality* has been a persistent focus of research and practice, forming the basis upon which reliable, valuable, and monetisable data-driven systems can be built.

In fact, *Data quality* has been a central topic in the literature with foundational frameworks emerging as early as the 1980s, such as Total Data Quality Management (TDQM) developed at MIT [100]. Subsequent initiatives include the WHO's Data Quality Assurance (DQA) framework [106], ISO 8000 [105], and ISO 25012 [104]. More recent comparative analyses, such as those by Miller et al. [97], provide extensive coverage of existing frameworks.

As illustrated in Figure 1, *data quality* constitutes a major sub-cluster of the internal processes perspective within our taxonomy, with strong links to other perspectives. Its importance is twofold:

- *Direct contribution to data value*: The *Quality & Utility* clusters, of which *data quality* is a core component, directly influence the valuation of datasets. Data quality shapes market price, monetisation potential, and internal prioritisation. Even highly promising datasets may fail to generate returns or meet strategic goals without sufficient quality.
- *Cross-perspective enabler:* Although situated under *Internal Processes*, data quality impacts all BSC perspectives. In the *Financial* perspective, it reduces valuation risk and underpins defensible pricing. For the *Customer* perspective, it ensures the delivery of accurate, relevant, and trustworthy insights that drive satisfaction and loyalty. Within *Learning & Growth*, it safeguards the reliability of innovation initiatives and analytical models, reducing rework and accelerating development.

In short, *data quality* provides the structural integrity of the taxonomy. It is the foundation that transforms raw information into a reliable strategic asset. Weak data quality undermines governance, compliance, operational efficiency, and advanced valuation techniques. Embedding it at the heart of the taxonomy ensures that data used across all strategic objectives is not only available but also fit for purpose.

A persistent challenge in this domain is the sheer number of proposed quality metrics, often with overlapping scope or definitions. Ballou et al. [9] identified 76 data quality metrics alone , many of which overlap or present only minor distinctions (e.g., timeliness vs. currency). In constructing our taxonomy, we consolidated several synonymous or closely related metrics into unified categories, where appropriate. For instance, we grouped clarity and understandability under a single concept, as both ultimately represent the same concept of preventing confusion when using data.

This consolidation process is, by nature, subjective and will likely evolve as the field matures. Miller et al. [98] and Schwabe et al. [128] similarly merged synonymous metrics within their frameworks. Nevertheless, inconsistencies across definitions and usage have slowed progress toward a unified standard and contributed to the proliferation of competing data quality frameworks [97]. These discrepancies hinder the development of a global data market where organisations can consistently assess both the value of their own data and the data they procure.

Figure 2 details the *data quality* sub-cluster. It is organised into four groups, each reflecting different dimensions of how well data meets technical, operational, and strategic requirements:

- *Fundamental* quality ensures that datasets are accurate, consistent, and valid, forming the non-negotiable baseline for any valuation or monetisation activity.
- *Contextual* quality adapts this baseline to domain-specific requirements such as stability, representativeness, and containment.
- *Resolution* addresses temporal properties such as timeliness and currency, which are crucial for markets and operational contexts where the value of data degrades rapidly.
- *Specialised* quality encompasses usability, clarity, and plausibility, ensuring that data assets are interpretable, fit-for-purpose, and actionable by end users.

## 3.1 Fundamental Metrics

Within the fundamental dimension, a variety of metrics are used to characterise and summarise datasets. Basic statistical descriptors such as mean, standard deviation, minimum, and maximum, belong to this group, though they are collectively referred to as *Statistics* in Figure 2 and are not discussed in detail here.

**Age** quantifies the time elapsed since a data item was created, updated, or last accessed [4, 54]. It is a critical indicator of timeliness and relevance, with two common forms:

- *Static age*: the interval since a dataset or record was first created, often guiding archival or tiered storage decisions.
- *Dynamic age*: the interval since the dataset was most recently updated or accessed, serving as a measure of freshness.

**Granularity** (also referred to as *Abundance* or *Data Frequency*) captures the level of detail at which data is collected, stored, and analysed. Fine-grained data enables highly specific insights but requires greater storage and processing capacity [88], while coarse-grained data reduces precision in favour of efficiency and trend detection [38]. The optimal resolution depends on analytical objectives, dataset characteristics, and the trade-off between precision and manageability [93].

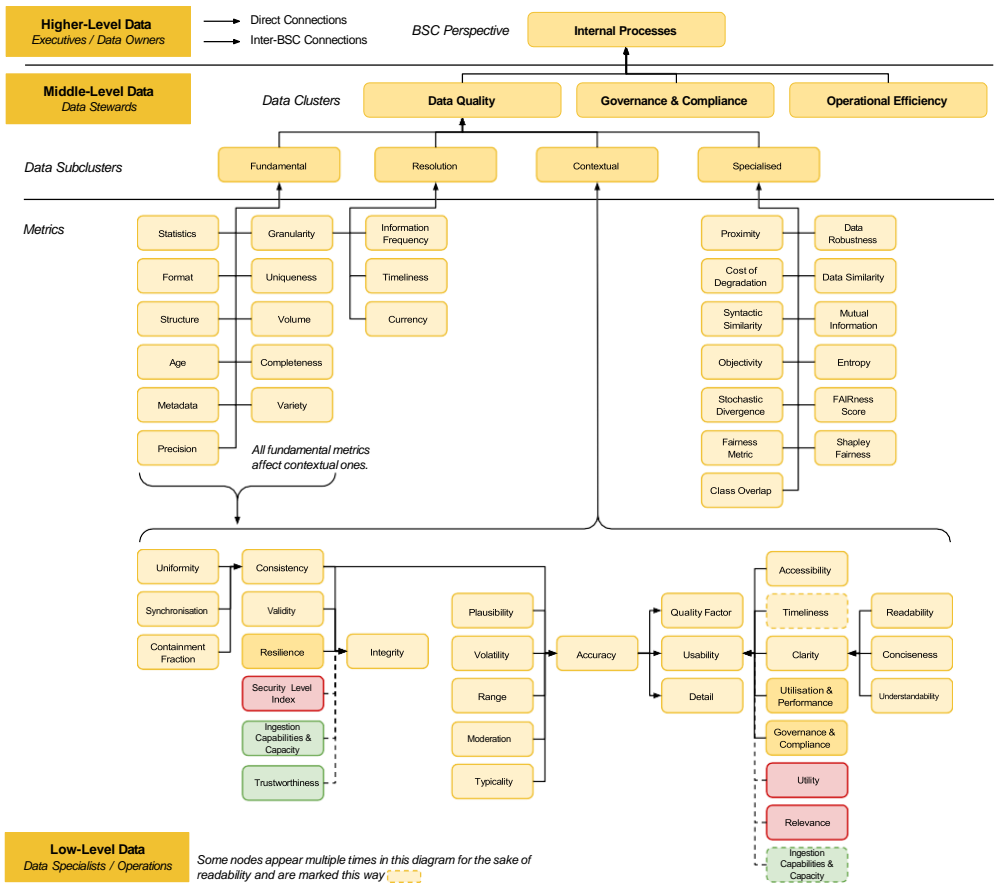

Fig. 2. Taxonomy of metrics and KPIs related to data quality. The fundamental data quality metrics affect all of the contextual data quality metrics.

Granularity is especially relevant in domains such as IoT and real-time analytics, where high-frequency updates (e.g., financial feeds or continuous sensor streams) are common [32, 38, 88, 93]. Although sometimes conflated with *velocity* [61], the two differ: velocity concerns the speed of data generation and transfer, whereas granularity denotes the temporal resolution of events.

**Precision** reflects the degree of fineness with which values are represented, directly influencing the reliability of analyses [10, 34, 141, 149]. It has three facets:

- *Numerical precision*: the number of significant digits or decimal places (e.g., 23.456 vs. 23.5).
- *Consistency*: the extent to which repeated measurements yield the same result under identical conditions.
- *Repeatability*: the ability to obtain the same measurement across multiple trials.

Precision applies beyond numeric data: for example, "Scarlet," "Azure," and "Emerald" provide greater specificity than the broad categories "Red," "Blue," or "Green."

**Uniqueness** (or, conversely, *Redundancy* in service contexts) measures the distinctiveness of dataset entries, ensuring that each entity is represented only once. This protects data integrity and consistency in decision-making [10, 51].

**Variety** (also termed *Multifacetedness*) denotes the heterogeneity of data across types, formats, and sources, including structured, semi-structured, and unstructured forms. Greater variety enhances analytical richness but increases integration and computational complexity [167].

**Volume** (alternatively referred to as *Quantity*, *Entries*, *Total Data Amount*, or *Number of Datasets Available*) captures the size of data available for analysis [20, 54, 67, 88, 151, 167]. It can be expressed as the number of records, points, or entries, influencing:

- *Statistical significance*, by reducing sampling error.
- *Pattern recognition*, by enabling discovery of correlations and trends.
- *Machine learning performance*, by providing sufficient training data.

Adequate dataset size ensures sufficient coverage for intended applications, while access to multiple datasets increases reuse and flexibility [20, 67].

**Metadata Compliance**, closely related to Profiling, provides descriptive information about data structure, provenance, content, and context. It supports interpretation, discovery, and usability, and is assessed in terms of semantic consistency, vocabulary adherence, and currency [57]. From a metric perspective, metadata compliance may be treated as a binary indicator (presence/absence of descriptive fields). Key functions include:

- *Discoverability*: enabling datasets to be searchable and interpretable (e.g., format specifications).
- *Quality assessment*: supporting evaluation of accuracy, completeness, and trustworthiness.
- *Context provision*: clarifying provenance, scope, and intended use.

**Format** (also known as *Format Compliance*, *Codification*, or *Available Formats*) measures the extent to which entries conform to expected patterns or structural requirements (e.g., geospatial coordinate formats) [32, 73, 149, 155].

**Structure** refers to the organisation of data—structured, semi-structured, or unstructured—and its syntactic and semantic consistency. Well-structured datasets enable interoperability and efficient analysis, while poorly structured ones impose costly pre-processing [10, 88, 155].

**Completeness** (sometimes termed *Appropriate Amount of Data*) indicates the degree to which all required or expected values are present. This widely studied metric [10, 20, 34, 51, 54, 61, 73, 149, 155] may involve verifying whether key fields are populated, or assessing coverage relative to a dataset's intended application.

## 3.2 Resolution Metrics

Within this cluster, we group metrics that are inherently time-dependent, dynamic, or strongly linked to data granularity. Among them, *Timeliness* [3, 9, 10, 49, 51, 59, 70, 80, 84, 97, 121, 125, 128, 131, 142, 158] and *Currency* [3, 9, 10, 17, 19, 32, 49, 51, 55, 59, 70, 84, 97, 98, 121, 125, 128, 131, 142, 158] are particularly well established.

**Timeliness** measures the delay between the occurrence of an event and the moment the corresponding data becomes available to the organisation. It is thus closely related to *Age*.

**Currency**, by contrast, captures whether data has lost value due to processing, modification, or the mere passage of time. It can be understood as the rate at which data value degrades, linking it conceptually to valuation techniques.

**Information Frequency (IF)** refers to the rate at which information is updated, accessed, or used within a system or process [141]. It describes the temporal dynamics of information flow, ensuring that information remains relevant and aligned with operational or decision-making needs. Although linked to operational efficiency, its conceptual overlap with quality justifies its inclusion in this cluster. IF can be further decomposed [141] into:

- *Raw Frequency Component (IFr)* – the baseline rate at which data becomes available.

- *True Frequency Component (IF)* – a refined measure that adjusts for external factors affecting information flow.
- *Frequency Tolerance (FT)* – the acceptable range of frequency deviations that still enables effective decision-making.
- *Node Frequency Requirement (FN)* – the minimum update rate required to satisfy the needs of a specific system or process.

This aggregation captures both quantitative and qualitative aspects of frequency, ensuring information availability with sufficient timeliness and precision to support decision-making processes [141]. While each component could be treated as an individual metric, their interdependence with IF led us to incorporate them collectively under this single construct.

## 3.3 Contextual Metrics

Contextual metrics evaluate data quality relative to a specific application, domain, or use case. They emphasise that quality is not absolute but depends on who uses the data, for what purpose, and under which conditions.

**Range** denotes the proportion of values falling within predefined minimum and maximum thresholds. It verifies whether data lies within expected operational boundaries and directly influences the *Utility* dimension. For numerical attributes, range checks ensure values remain valid according to domain-specific or statistically derived limits (e.g., extrema or quartile ranges) [59, 95].

**Moderation** measures the share of data points that fall within a given confidence interval. For instance, in a normal distribution, roughly 99.7% of data should fall within three standard deviations of the mean. This reflects dataset stability and reliability by testing adherence to high-confidence expectations [95].

**Typicality** captures how representative a single item is relative to established patterns in a dataset. It highlights conformity to expected norms and is useful for identifying anomalies or unusual events, particularly in dynamic environments [95].

**Volatility** quantifies how often and, in some cases by how much, data values change over time [10]. Estimation methods vary depending on application, but volatility consistently reflects instability and update frequency. In finance, for example, it also measures the magnitude of fluctuations.

**Consistency** assesses the alignment of data with predefined rules and standards, making it closely related to *Veracity* and *Reliability* [167]. Constraints may be structural (e.g., database rules) or semantic (e.g., organisational standards). Following [6], intra-consistency evaluates whether tuples satisfy association rules, while inter-consistency ensures uniformity across datasets, systems, or applications. Core aspects of consistency include [10, 20, 34, 59, 73, 88, 95]:

- *Integrity* – Accuracy and dependability regardless of access point or processing route.
- *Uniformity* – Standardisation of formats, naming conventions, and measurement units.
- *Synchronisation* – Ensuring updates propagate across systems without mismatch.
- *Validation* – Rule-based enforcement of types, ranges, and relationships.
- *Error prevention* – Detection and correction of conflicting entries.
- *Data governance* – Organisational standards and policies.

**Containment Fraction** evaluates how much of one dataset overlaps with another, supporting redundancy detection, consistency checks, and storage optimisation in distributed systems [133].

**Integrity** represents correctness, reliability, and compliance with standards. It guarantees completeness, authenticity, and fitness for use [4, 10, 32, 34, 51, 59, 67, 155]. *Integrity* strongly depends on *Consistency* and *Validity* but also on factors from other clusters such as the *Security Level Index* which attributes a score based on how safely and securely data is stored, the *Ingestion Capabilities & Capacity* of the system in terms of data, and the overall *Trustworthiness* of the system.

**Uniformity** reinforces consistency by requiring identical data representations (e.g., date formats), without evaluating logical correctness, which falls under validity.

**Validity** ensures individual values adhere to logical and operational rules (e.g., ages or costs must be positive) [34, 59, 67]. As an analogy, in a puzzle set, *integrity* ensures all original pieces are present and undamaged; *validity* confirms each piece is correctly shaped; *consistency* ensures the pieces fit together without conflict; *uniformity* guarantees all pieces share the same material and finish.

**Accuracy** captures the extent to which data reflects the true or intended state of the world. It combines correctness with decision-making utility [10, 20, 32, 51, 59, 61, 67, 141, 149, 155]. Contributing metrics include range, consistency, moderation, and typicality [95]. Thresholds are application-specific, with tolerance for deviations depending on context.

**Quality Factor**, proposed by [141], links quality to business innovation by combining accuracy with frequency. The frequency component corresponds to *Information Frequency* (IF), discussed under resolution metrics.

**Detail** measures whether information is recorded at a sufficient level of precision, linking accuracy directly to data valuation [51].

**Plausibility** evaluates whether data aligns with real-world knowledge. The European Central Bank framework extends this to anomaly detection and outlier handling [10, 20, 34, 51, 57, 123, 155]. As a KPI, plausibility can combine range, consistency, domain-specific rules, and anomaly detection.

**Usability** is widely discussed but context-sensitive. ISO 9241:11 (2018) defines it as "a benchmarking tool that can be used to determine the extent to which a system, product, or service can be used by specific users to achieve the goals determined by the effectiveness, efficiency, and satisfaction of its users." [57]. Building on this, and integrating insights from [10, 32, 34, 54, 149, 155], we expand usability to mean: "The ease and efficiency with which quality data, defined by *Data-Value*, *Accuracy*, *Integrity*, and *Completeness*, can be accessed (through *Communication*, *Accessibility*, and *Timeliness*), understood (*Clarity*), and effectively used (*Ease-of-use*, linked to *Operational Efficiency*, *Openness & Performance*, *Relevance*, or *Utility*) by users to complete specific tasks" Because it aggregates multiple components, usability is best treated as a KPI rather than a single metric.

**Accessibility** refers to the ease with which users can access and utilise data [42, 98]. It is often associated to discoverability, but also to *Format* based on usability and *Metadata* based on available information and searchability.

**Clarity** is a composite KPI reflecting how unambiguously and readably information is presented, improving interpretability and usability [10, 20, 34, 51, 57, 73]. Since *Clarity* is a combination of factors, it can be treated as a KPI where the values are agglomerated and weighted. *Clarity* serves as an integrative measure reflecting several metrics (see the following ones), emphasising the presentation of high-quality, actionable, and user-centred information across diverse applications, from open-data platforms to governance and compliance systems.

**Conciseness** assesses the brevity and directness of data presentation, minimising redundancy while retaining meaning. It reduces ambiguity and enhances clarity by removing irrelevant detail [10, 51].

**Understandability** (or *Ease of Understanding*) evaluates how easily users can interpret information. It is supported by clear field names, precise definitions, and unambiguous units, enabling non-experts to work with datasets [10, 57, 73].

**Readability** enhances clarity by ensuring data is structured and presented for intuitive use. This includes formatting, user interface design, and visualisation, all of which improve interaction and interpretation [9, 107].

## 3.4 Specialised Metrics

Specialised metrics extend beyond general-purpose quality dimensions (such as accuracy or completeness) and are tailored to particular domains, data types, or regulatory contexts. They often address industry-specific or task-specific requirements, providing nuanced insights into data quality and its role in valuation or monetisation.

**Entropy** and related measures—including *Mutual Information*, *Information Content (IC)*, and Shapley-value–based expressions (e.g., *Shapley Fairness*, *Shapley Robust*)—provide advanced characterisations of datasets' structure and information content. *Entropy*, rooted in information theory, measures uncertainty, randomness, or heterogeneity within a dataset [61, 151]. It is widely used to evaluate informativeness, uncertainty, and representational effectiveness. Although often applied in valuation, we include it here as a quality metric due to its strong connection with information richness. Variants such as Shannon's entropy, heterogeneity indices, joint entropy, and information scores are context-dependent, and their use depends on the chosen analytical strategy.

**Mutual Information** quantifies the amount of information shared between two random variables, measuring how much knowledge of one reduces uncertainty about the other. Mathematically, it is expressed as the Kullback–Leibler divergence between their joint distribution and the product of their marginals. This makes it a powerful tool for detecting dependencies in datasets, with applications in data science, machine learning, and signal processing [4, 31]. It supports the assessment of relevance, consistency, and completeness, highlights incoherence or gaps in data, and is a key metric for feature selection and dimensionality reduction.

**Shapley Fairness and Fairness Metrics**, grounded in cooperative game theory, extend the concept of *Shapley Values* to quantify the contribution of individual data points or datasets to predictive model performance [2]. Shapley-based approaches ensure that allocations respect fairness principles such as balance, symmetry, and additivity. They can reveal redundancy or noise by identifying low-contributing data, thus improving dataset quality. For instance, fairness metrics have been applied to cache space allocation in virtual machines, ensuring proportional resource sharing based on weights [65, 102].

**FAIRness Score** (distinct from fairness metrics) evaluates compliance with the FAIR principles (Findability, Accessibility, Interoperability, and Reusability). Tools such as CkanFAIR automatically compute FAIRness scores to assess dataset readiness for sharing and reuse [86].

**Data Similarity**, as defined in this work, encompasses measures such as Euclidean distance, cosine similarity, Jaccard index, Kolmogorov–Smirnov tests, Mann–Whitney tests, and Levene statistics [4, 39, 151]. These measures assess similarity or divergence across sets, vectors, or distributions; for instance, the Jaccard index captures dataset overlap.

**Syntactic Similarity** evaluates how closely values align in terms of syntax and representation, using methods such as Levenshtein distance, edit distance, q-gram analysis, or MinHash-based overlap [151]. These are widely used in text and string-matching tasks.

**Stochastic Divergence**, introduced in this work, captures similarity between probability distributions using measures such as t-tests, Jensen–Shannon divergence, Wasserstein distance, or identity-based exact matches [33, 79, 96].

**Objectivity** measures the degree to which data is unbiased and free from subjective influence during collection, evaluation, or use. It is an intrinsic dimension of quality, essential for reliability and credibility in decision-making [10, 20, 34]. Metrics associated with bias detection such as statistical parity, distributional skewness, and equalised odds are directly linked to this concept.

**Cost of Degradation** quantifies quality loss resulting from transformations, for example in privacy-preserving techniques that introduce distortion [139].

**Proximity** measures the physical distance between a data source (e.g., a sensor) and the event it observes, relevant in contexts where measurement accuracy depends on location [4].

**Data Robustness** reflects the resilience of datasets to perturbations, faults, or shifting contexts. Robust data remains valuable and usable across diverse prediction tasks or operational environments. As emphasised by [2], robustness is particularly important in the context of emerging data marketplaces, where resilience ensures long-term utility and value preservation.

**Class Overlap** (also referred to as *Class Similarity*) measures the extent to which classes or clusters in a dataset intersect, based on regions where data points are associated with multiple labels. This metric provides insight into class separability and dataset complexity [56, 124, 138].

In summary, metrics within the *Data Quality* cluster form the foundation for reliable data valuation and monetisation practices. By ensuring that data is accurate, consistent, complete, and contextually relevant, organisations enhance trust in their analytical outputs and strategic decisions. High data quality not only improves internal efficiency but also strengthens the market value of data assets, enabling their reuse, exchange, and commercial exploitation under reliable and transparent conditions.

## 4 DATA GOVERNANCE & COMPLIANCE

This cluster encompasses indicators and metrics that evaluate adherence to data governance principles and regulatory standards, including privacy and security requirements. By ensuring alignment with legal and ethical guidelines, these metrics mitigate the risks of breaches and regulatory non-compliance, thereby fostering trust among clients and stakeholders.

*Compliance* refers to metrics and KPIs that track conformity with legal, regulatory, and standards-based requirements, such as ISO 20022. It covers both internal and external oversight mechanisms. As illustrated in Figure 1, compliance is pervasive, influencing every cluster and sub-cluster, and thus underpins all aspects of an organisation's operations.

*Data governance* encompasses the processes, methods, roles, policies, standards, and metrics that ensure proper management and responsible use of data across its lifecycle, from discovery and collection to processing, storage, analysis, and disposal [94]. Effective governance promotes accuracy, security, consistency, and accessibility, while ensuring compliance and supporting organisational objectives.

It is important to note that governance extends far beyond the scope of the specific metrics described here, as it is inherently linked to strategy, security, and business architecture. In this taxonomy, we use the *Governance* cluster primarily as a structural layer that connects compliance, safety, and security. Accordingly, we categorise governance-related metrics and considerations into three sub-clusters: *Compliance*, *Openness & Ownership*, and *Risks, Safety, & Security*. Figure 3 illustrates the *Governance & Compliance* cluster and its sub-clusters.

### 4.1 Compliance

Compliance involves metrics designed to track adherence to legal, standards, and regulatory requirements. Given the broadness of this aspect, they were further subdivided in two cases:

- *External Compliance*, in other words, legal and regulatory requirements and;
- *Internal Compliance* regulated by standards, policies and other definitions that facilitate handling of data within the organisation.

These types of metrics are often context-dependent due to the regulations and policies nuances, independent of standardised models across industries. Furthermore, aggravating the discrepancy in the data monetisation environment, the maturity level in the domain is too low for the existence of broad standards, metrics, and policies.

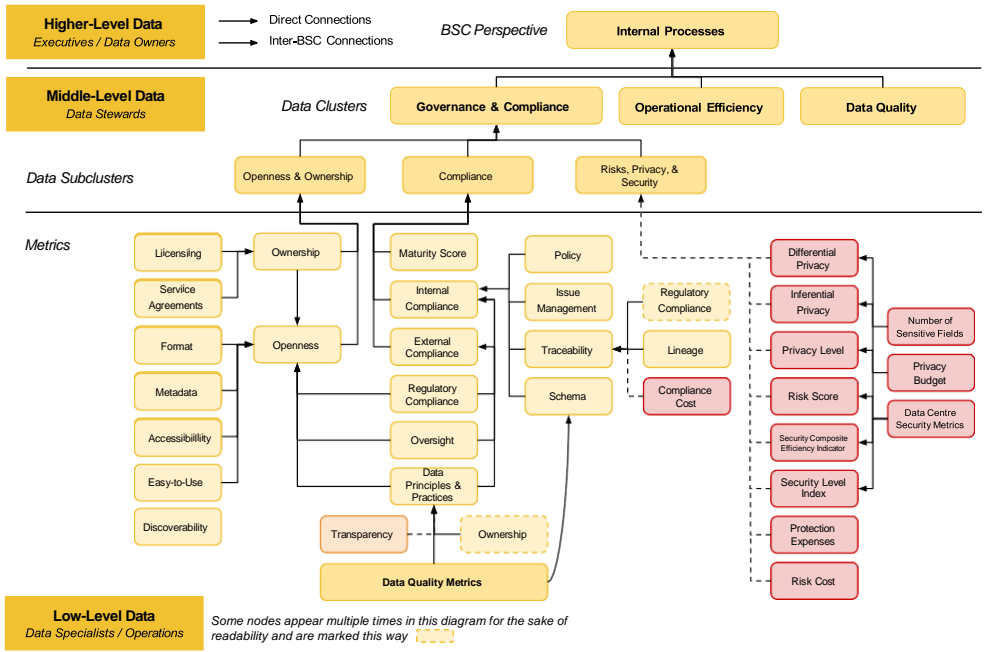

Fig. 3. Taxonomy of metrics and KPIs concerning data governance and compliance. Data quality metrics directly affect *Data Principles & Practices*, and *Schema*. Metrics grouped under risks, privacy, & security span both the *Governance & Compliance* and the *Financial* clusters.

**Maturity Score:** The evaluations in [161] and [94] address gaps in the literature by identifying key dimensions of data governance assessment, including *Policy*, *Oversight/Audit*, *Data Principles and Practices* (and their correct application), *Maturity Score* (from maturity assessments), and *Data Issue Management*. Together, these elements provide a structured framework for analysing and improving governance practices. For instance, the DAMA-DMBOK and COBIT-based approach described in [161] enables a comprehensive evaluation of governance maturity and offers actionable insights for organisations seeking to enhance governance capabilities.

Metrics play a critical role in informing maturity models. Well-defined indicators provide the evidence needed to measure progress, highlight areas for improvement, and verify alignment with compliance and organisational goals. In this way, metrics not only support governance monitoring but also directly contribute to the estimation and refinement of maturity scores. For example, the Capability Maturity Model (CMM) has been used as an indirect valuation technique: by evaluating data processes along dimensions such as cost, quality, and utility, CMM-based frameworks enhance both data governance and data value [18, 48].

**Internal and External Compliance** - Internal compliance metrics focus on and assess the conformity of data practices with established internal standards and policies while external compliance includes indicators for legal compliance, compliance costs (from the data valuation cluster), licensing adherence, and the presence of data-sharing policies. Both integrate *Regulatory Compliance*, *Oversight* (or *Audit*), and *Data Principles & Practices*.

**Regulatory Compliance** (linked to *Compliance Cost*) measures adherence to legal requirements and regulatory frameworks, thereby reducing risks associated with non-compliance. Metrics such as *Compliance Cost* and *Risk Score* are integral to evaluating financial data models. For example, compliance with standards such as ISO 20022 ensures regulatory alignment in electronic data exchanges within the financial sector, streamlining international operations and mitigating compliance risks [127]. In cybersecurity, compliance metrics assess whether data protection practices meet national security regulations, highlighting their role in safeguarding operational security and legal conformity [72].

Ownership-related metrics, often identified in data integration efforts, further emphasise the importance of responsibility and stewardship. This underscores the context-dependent nature of compliance metrics, which must adapt to organisational structures and governance frameworks [161].

The growing prominence of compliance reflects heightened regulatory scrutiny of organisations handling large volumes of sensitive data. As noted by [53, 136], indicators such as *Compliance Cost* and *Licensing* are essential for aligning data practices with evolving legal standards. These studies also highlight the challenges faced by data owners in navigating complex legal environments and the significant financial implications of maintaining compliance.

**Oversight** (also referred to as *Audit*) – Although explicit *Oversight* metrics are rarely defined in the literature, evaluating an organisation's ability to monitor and validate compliance with governance standards, processes, and policies remains essential. This includes establishing policies for regular audits to verify data integrity and ensure effective implementation of governance practices [161].

**Data Principles & Practices** (also referred to as *Data Standards* or *Standardisation*) concern the secure, efficient, and cost-effective discovery, collection, processing, analysis, storage, and defensible disposal of large volumes and high-velocity streams of structured and unstructured data. The adoption of standards and policies at early stages can ensure proper implementation of these processes.

Although standardisation is frequently emphasised in the literature, it is not consistently associated with explicit metrics. For example, several works [59, 155, 161] highlight its importance within evaluation frameworks, yet often without specifying concrete measurement approaches. In this taxonomy, we therefore group these contributions as indicative considerations, acknowledging their evaluative role despite the lack of standardised metrics.

**Transparency** is a multifaceted metric that affects both data and data processes. It ensures that data is measurable, verifiable, and interpretable, thereby supporting quality, governance, and model fairness [73, 102, 161]. Transparency is also closely tied to *Reputation*, particularly through its role in service-level agreements (SLAs) and business processes, where it guarantees that data is measured and reported according to predefined procedures.

As a governance consideration, transparency requires that data sources, transformations, and decision-making criteria are openly documented and accessible. This openness enables stakeholders to audit and validate data use, reducing bias and fostering trust in both data and derived insights.

**Policy** refers to a set of formalised principles, rules, or guidelines established by organisations, institutions, or governments to regulate internal or operational practices [32, 82]. Policies are typically designed to ensure compliance with legal requirements while incorporating organisation-specific strategies or controls. For example, [82] defines policies that meet GDPR compliance but also introduce stricter internal safeguards for sensitive information.

As a metric, policy evaluation is closely related to *Data Principles & Practices*. Possible measures include adherence rates, audit results, survey responses, and stakeholder feedback, which together indicate the extent to which policies are effectively implemented and followed.

**Issue Management** refers to the processes established for identifying, tracking, and resolving data-related problems, particularly those affecting compliance. For example, [161] discusses *Data Issue Management* in the context of corporate data glossaries, where inconsistent or erroneous definitions may create integration challenges across organisational services. Effective issue management ensures that such problems are systematically addressed to maintain data quality, consistency, and compliance.

**Traceability** refers to the ability to track and verify data across its entire lifecycle. It essentially refers to the "who", "where", and "when". It encompasses *Data Lineage*—the flow of data from its source through transformations to its final use—and *Provenance*, which records the origin, history, and modifications of data, including who made changes, when, and what was altered [128]. Maintaining audit trails ensures accountability and transparency, while also supporting compliance with regulatory and governance standards. Traceability further includes *Addressability*, the ability to identify and, if necessary, contact or reference the original data source.

Metrics commonly associated with traceability include:

- *Data Lineage Completeness* – percentage of data elements with documented lineage.
- *Provenance Documentation (PD)* – percentage of data elements with provenance records.
- *Audit Trail Coverage* – percentage of changes captured in an audit trail.
- *Compliance Rate* – percentage of processes adhering to governance and regulatory standards.

Additional traceability-related measures have been derived from the Agent–Goal–Decision–Information (AGDI) model [76, 137]. However, these are often context-specific, and their applicability depends on the organisational or domain setting. For this taxonomy, we group such measures under the broader category of *Traceability*, leaving readers to determine the relevance of general-purpose or decision-information model–based metrics to their own context.

**Lineage** captures the origin, transformations, and overall history of data—in essence, describing the "what" and "how". It can be established as a requirement within both internal and external compliance frameworks, often embedded in schema definitions, metadata, and related compliance metrics. In this work, lineage is treated as a metric that reflects the presence (or absence) of provenance information.

While *traceability* metrics are directly applicable to operational data management, particularly within decision support systems, lineage is a broader concept. It encompasses the ability to link and follow relationships not only between data elements but also across processes, decisions, and goals within a system.

**Schema**, as a metric, evaluates compliance with predefined data structures. It considers the presence or absence of desired attributes such as *Clarity* of definitions, *Comprehensiveness*, *Flexibility*, *Robustness*, *Domain Precision*, *Identifiability*, *Obtainability*, *Relevance*, *Adaptability* to semi-structured data, semantic alignment, standardisation for interoperability, and support for integration processes [10, 73, 131].

As discussed in [22], several metrics have been proposed to assess the quality of data models across conceptual, logical, and physical levels. These requirements are typically ratio-based and intrinsically tied to schema design. For instance, NFT measures the number of fact tables that comply with the schema. Other examples include NDT, NSDT, and NAFT. In this work, we group such measures collectively under the *Schema* category.

## 4.2 Openness & Ownership

This cluster covers metrics related to data openness and ownership, reflecting both the accessibility of data and the rights associated with its use. While openness emphasises transparency and sharing, ownership focuses on control, licensing, and service agreements. Together, these dimensions are critical for data governance, valuation, and compliance.

**Openness** measures the degree to which datasets are accessible, usable, and reusable by the public, subject to intended use and regulatory obligations. It includes the provision of machine-readable formats, non-proprietary standards, and open licenses that allow free use, redistribution, and modification [18, 75, 84, 116, 155]. As a metric, openness incorporates principles of *Accessibility* (linked to the *Operational Efficiency* cluster), *Interoperability*, and open *Licensing*. The 5-star Model of Openness [132] is often applied to estimate this metric. Depending on context, factors such as *Ease-of-use* may also be considered.

**Ownership** represents the level of control and rights an organisation holds over a dataset, ranging from full ownership to restrictive licensing or dependence on third-party services [47]. Ownership affects multiple valuation factors, including *Privacy*, and *Usability*, and influences integration, access, and flexibility [32, 47]. While higher ownership often requires greater upfront investment, it reduces long-term licensing costs and dependency risks.

**Licensing** is a context-dependent factor shaped by industry regulations and national legal frameworks [3, 17, 32]. As shown in Figure 3, licensing directly affects ownership: open licenses and internally managed datasets maximise ownership value, while restrictive licenses reduce organisational control. Because outright ownership can be viewed as the inverse of licensing restrictions, we consider *Licensing* as the primary measurable metric.

**Service Agreements** refer to formal contracts that define data ownership, licensing terms, compliance obligations, and usage rights. As a metric, they are evaluated in terms of clarity, regulatory alignment, and flexibility, all of which influence data utility and accessibility.

### 4.3 Risks, Privacy, and Security

This cluster includes metrics that evaluate risks associated with data management, system security, and privacy. They are used to assess the effectiveness of protection measures and apply broadly across industries. While these metrics are grouped under the *Data Valuation* cluster in our taxonomy, they are also considered part of the *Governance & Compliance* cluster.

**Risk Cost** (also referred to as *Regulatory Risk Index*) accounts for the financial and reputational impacts of data loss, compromise, or misuse [19, 82]. It includes direct costs, such as regulatory fines and data replacement, as well as indirect consequences such as reputational damage or competitive disadvantage.

**Protection Expense** measures the cost of implementing safeguards (e.g., encryption, access control) and quantifies the financial impact of potential breaches. Its value depends on the legal environment, data type, and organisational risk profile. These metrics support privacy and security management by evaluating the ability to ensure integrity, prevent unauthorised access, and maintain compliance. Applications span multiple domains, including IoT [38], federated learning [146], data marketplaces [87], and incentive mechanisms balancing privacy and participation [159].

**Risk Score** is a quantified indicator of risk exposure, derived from factors such as data sensitivity, compliance obligations, breach likelihood, and potential impact. It supports prioritisation in protection strategies and informs decision-making in risk management [82, 88].

**Differential Privacy** provides a formal privacy guarantee by bounding how much the output of an algorithm can change when a single individual's data is added or removed [38, 87, 146, 159]. It is widely applied to assess modifications in datasets and protect against re-identification. Related metrics include *Inferential Privacy*, which supports the estimation of differential privacy levels [38].

**Privacy Budget** quantifies the amount of privacy "spent" by a data owner in exchange for participation in marketplaces or collaborative models. It determines the level of noise introduced into data or gradients, balancing privacy protection and utility [53, 146]. As shown in Figure 3, it is hierarchically related to *Differential Privacy*, alongside metrics such as *Number of Sensitive Fields* [82] and *Data Centre Security Metrics*.

**Privacy Level** provides a quantitative measure of data protection, often expressed by the differential privacy parameter $\epsilon$. Lower values indicate stronger privacy guarantees but reduced utility, while higher values relax privacy for improved usability [45, 87, 157, 159]. Closely linked is privacy sensitivity, reflecting the value individuals or organisations place on privacy, which influences both the required level of protection and compensation mechanisms in data marketplaces [87, 157].

Case studies highlight how privacy-preserving methods are applied in practice. For example, [157] describes the use of differential privacy and synthetic data to safeguard sensitive governmental datasets in New Zealand, balancing utility with strict privacy requirements. This context-specific approach reflects the reliance on privacy techniques tailored to the needs of the data being handled.

**Security Composite Efficiency Indicator (SCEI)** is a holistic index of Information Security and Cybersecurity System (ISCSS) performance [72]. It is calculated as the mean of five sub-components:

- *Equipping Coefficient with Cyber Defence Means* – system preparedness against cyberattacks.
- *Technical Readiness Coefficient* – operational readiness of security mechanisms.
- *Equipping Coefficient with Serviceable Defence Means* – availability of functional cybersecurity tools.
- *Staffing Coefficient with IT Administrators* – adequacy of IT system administration resources.
- *Staffing Coefficient with Service Personnel* – sufficiency of service staff for security operations.

**Security Level Index (SLI)** quantifies the effectiveness of security mechanisms, incorporating indicators such as encryption strength, access control, incident response capacity, and compliance with standards [28, 72]. SLI is closely tied to access security and assesses resilience to unauthorised use. For data centres specifically, a wide range of operational metrics have been proposed, including *Average Comparisons per Rule*, *Application Transaction Rate*, *Concurrent Connections*, *Defence Depth*, *Firewall Complexity*, and *Vulnerability Exposure* [119]. We group these under *Data Centre Security Metrics*, which can feed into the calculation of an SLI.

SLI plays a critical role in hierarchical systems by stratifying protection levels according to data sensitivity and operational requirements, optimising resource allocation and risk mitigation. In IoT environments, trust frameworks are often integrated to evaluate reliability and detect malicious behaviour [20, 71]. More broadly, SLI supports large-scale systems in addressing challenges related to data volume, velocity, and system resilience [63, 167].

In conclusion, the *Governance & Compliance* cluster establishes the structural and regulatory backbone for data valuation and monetisation. By ensuring that data practices align with ethical, legal, and organisational standards, governance and compliance metrics mitigate risks, foster accountability, and reinforce stakeholder trust. These elements collectively enable the sustainable exploitation of data assets, ensuring that value generation is both responsible and resilient to evolving regulatory environments.

## 5 OPERATIONAL EFFICIENCY

Several key performance metrics in computing, data management, and networking can be attributed to either *Technology & Infrastructure* or *Operational Efficiency*, depending on context. While *Technology & Infrastructure* (to be detailed in future work) concerns the hardware, network, and system architecture that define technical capacity and constraints, *Operational Efficiency* focuses on how effectively those resources are managed, monitored, and optimised to maximise performance. Applications range from customer relationship management (CRM) analytics to high-speed railway data systems [25, 39, 145, 150]. Metrics such as *Latency*, *Bandwidth*, *Network Traffic Overhead*, and *Throughput* span both major clusters.

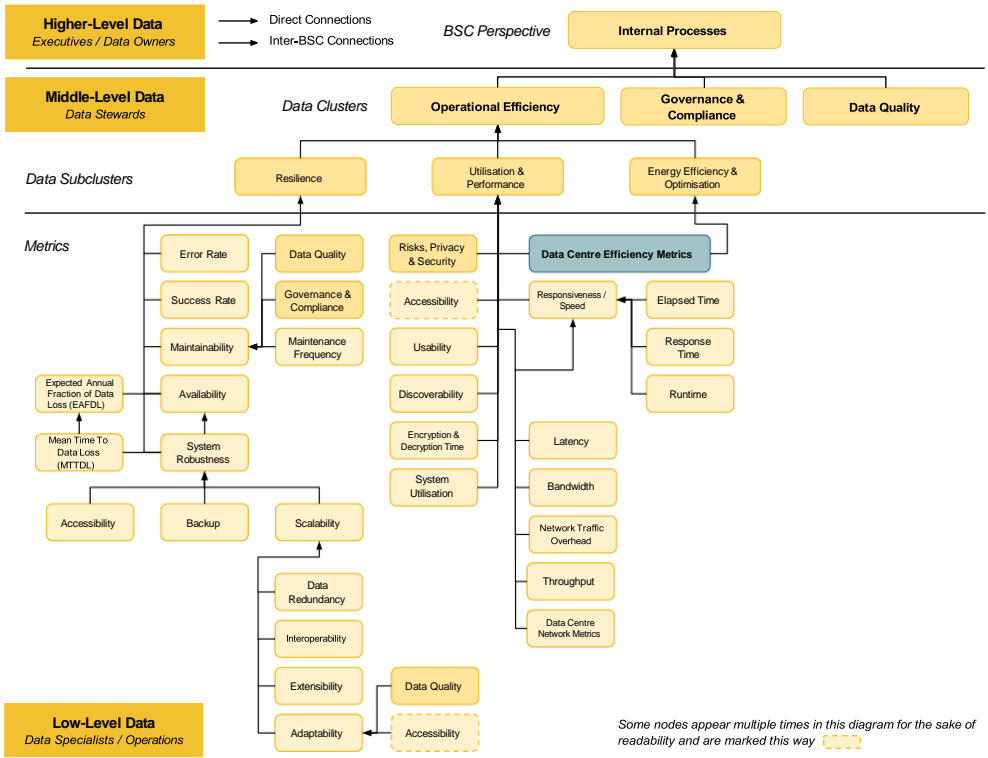

Fig. 4. Taxonomy of metrics and KPIs concerning data quality. Each sub-cluster captures a specific dimension of performance management.

Following an extensive literature review, we classified metrics and KPIs relevant to operational efficiency into three principal areas: **Resilience**, **Utilisation & Performance**, and **Energy Efficiency & Optimisation**. The analysis focuses on explicitly documented metrics and highlights context dependencies where applicable.

## 5.1 Resilience

Resilience-related metrics evaluate the reliability and robustness of data systems, ensuring continuous access to resources regardless of hosting environment, whether cloud-based or on-premises [18, 25, 143]. These metrics address *System Robustness*, *Availability* (linked to *Retrievability*) [14, 32, 51, 63, 67, 75, 89, 97, 98, 108, 116, 117, 121, 131, 171], *Error Rates* (including *Error Ratio*, *Error Count*, and *Inter-Server Error Rate (ISER)*) [14, 18, 46, 48, 53, 64, 67, 143, 153], *Success Rate*, and *Maintainability* [10, 149].

**Error Rate and Success Rate** quantify system reliability and operational effectiveness. *Error Rate* measures the proportion of failed operations (per transaction, time, or request), serving as a KPI for service stability and troubleshooting [64, 143]. Conversely, *Success Rate* and *Task Completion Rate* indicate the proportion of successfully completed processes and are widely used as indicators of system efficiency and user satisfaction [32, 57, 88, 149].

**Maintainability** reflects the ease and frequency of maintenance operations, depending on governance and quality practices [10, 149]. Related metrics such as *Maintenance Frequency* underscore the importance of regular checks to minimise downtime and maintain system stability [133].

**Mean Time to Data Loss (MTTDL)** and **Expected Annual Fraction of Data Loss (EAFDL)** estimate the expected duration or probability before a storage system experiences unrecoverable data loss. As defined in [66], MTTDL and EAFDL are critical metrics used to evaluate the reliability of storage systems.

**Availability** (or *Retrievability*) measures the extent to which data can be accessed and retrieved without error or restriction [14, 32, 51, 63, 67, 75, 89, 97, 98, 108, 116, 117, 121, 131, 171]. Additionally, it encompasses the readiness and openness of data to be used by the public or other stakeholders [32]. While availability ensures data existence and readiness, retrievability focuses on the actual ability to fetch and use the data effectively without technical or administrative barriers. Both are interlinked and treated as a unified metric. Availability can also serve as a composite KPI when combined with others metrics, for instance, securing data to its intended use also implies secure authorised users access and management of the data [116].

**System Robustness** evaluates replication integrity, equitable data distribution, and fault tolerance [2, 63, 81]. It also encompasses supporting metrics such as *Accessibility*, *Backup*, and *Scalability*.

**Accessibility**, in the context of data and information systems, refers to the ease with which users can access and utilise data. It impacts several key aspects:

- *Availability*: Data should be readily available to authorised users when needed, without unnecessary delays or barriers.
- *Usability*: Data should be presented in a format that is easy to understand and navigate, allowing users to efficiently find and use the information they need.
- *Interoperability*: Data should be compatible with various systems and applications, enabling seamless integration and exchange of information across different platforms.
- *Inclusivity*: Accessibility also considers the needs of diverse user groups, including those with disabilities. This means ensuring that data and information systems are designed to be usable by individuals with varying abilities, such as providing alternative formats or assistive technologies.
- *Security and Permissions*: While ensuring accessibility, it is also important to manage access controls and permissions to protect sensitive data from unauthorised access.
- *Open to use / licensing*: Data should be licensed in a clear manner for users.
- *Format*: Follow prestructured definitions for the data content structure, for example, does the dataset match Berners-Lee's "linked data principles".

Overall, *Accessibility* is crucial for maximising the value of data, as it ensures that users can effectively leverage the information for decision-making, analysis, and other purposes. It measures how accessible the platform is for users in terms of ease of navigation, search, and data retrieval. Chu et al [32] offer a way of measuring accessibility as a composite KPI for the interested reader.

**Backup**, based on [63], is defined as a critical component of the broader system - backup and recovery capabilities – which focuses on the system's ability to preserve data securely and ensure its availability during recovery scenarios. In this taxonomy, we group both backup and recovery capabilities under a single metric, as both are essential for maintaining operational continuity.

**Scalability** refers to the ability of a system, framework, or architecture to handle increased workloads or growing amounts of data by adding resources (e.g., hardware, software) or optimising existing ones, without compromising performance or efficiency [114, 133, 148]. *System Concurrent Processing Capability* refers to a system's ability to handle multiple processes simultaneously. *Scalability* is directly linked to this capability, as a scalable system must be able to manage increased concurrency [63]. Another linked concept (grouped under *Scalability*) is *Elasticity*. *Elasticity* is the ability of a system to dynamically adapt to changing workloads by provisioning or removing resources in real time [89, 148]. Numerous metrics can be linked to scalability. For example the system throughput refers to the amount of data processed, or the number of tasks completed within a specific time frame. In IoT systems, improvements in both throughput and latency reflect the system's ability to handle increasing data loads and user demands without performance degradation [114, 148]. Other factors that help to measure *Scalability* are *Data Redundancy*, *Interoperability*, *Extensibility*, and *Adaptability*.

**Data Redundancy** reduction, efficient deduplication, and compression techniques, widely adopted in large-scale storage systems like cloud platforms, have been shown to significantly enhance overall data quality and system performance [140, 148]). *Data Redundancy* has been grouped under *Uniqueness* in Appendix C.

**Interoperability** (linked to *Compatibility*, *Integration Capabilities*, and *Implementability*) defines how effectively systems and datasets communicate and exchange data. It depends on syntactic and semantic alignment and directly influences data valuation [10, 88].

**Extensibility** measures the capacity to add new components or functionalities without major redesign [43, 148].

**Adaptability**, also referred to as *Flexibility* or *Versatility*, represents the ability of a system (e.g. infrastructure and IoT [148, 155]), process (e.g. adaptability evaluation using Pettigrew's Context-Content-Process (CCP) model [149]), or service (e.g. data model [127]) to efficiently adjust to evolving conditions, changing requirements, or dynamic environments. It ensures systems remain effective across various contexts by allowing seamless integration [57, 140, 148].

## 5.2 Utilisation & Performance

As already mentioned, this cluster focuses on metrics that evaluate how efficiently system resources are used rather than the capacity available.

**System Utilisation** aggregates metrics for CPU, memory, storage, and network usage, providing insight into current system load and resource distribution [41, 60, 64, 119, 143, 150].

**Encryption & Decryption Time** evaluate the computational efficiency of privacy-preserving mechanisms [39]. They measure the processing overhead of encryption tasks to ensure that security measures do not compromise operational performance. Related indicators, such as the number of weak logins or detection accuracy, are included under *Data Centre Security Metrics*.

**Responsiveness** measures system reactivity and user experience, particularly in time-sensitive environments like e-commerce or logistics. Response delays beyond one second can degrade perceived efficiency, while those exceeding ten seconds often result in user abandonment [170]. Optimising responsiveness involves balancing network traffic, latency, and throughput through methods such as dynamic load balancing, adaptive caching, and predictive prefetching [5, 65, 130].

**Elapsed Time** measures total process or transaction duration, incorporating all delays due to computation, contention, and communication [115]. Efficient workload distribution and caching can substantially reduce elapsed time [65, 115, 130, 150]. This metric is particularly relevant in cloud computing, where system workloads fluctuate dynamically, affecting transaction speeds and response predictability.

**Response Time** quantifies the duration between request and response, often expressed as average or percentile-based values (e.g., 95th percentile) to reflect worst-case scenarios [150]. This metric is key in applications requiring real-time interactions, such as financial transactions, machine learning inference pipelines, and high-frequency trading platforms.

**Runtime** captures total execution duration from start to completion, including processing, memory access, and I/O. It directly impacts responsiveness and overall system throughput [130, 140].

**Latency, Bandwidth, Network Traffic Overhead, and Throughput** collectively assess system speed and communication efficiency. Latency metrics (e.g., end-to-end, inter-server, uplink/-downlink) are essential in real-time applications such as streaming, IoT, and cloud computing [4, 5, 25, 60, 64, 68, 114, 115, 120, 130, 153]. Techniques like data filtering, traffic optimisation, and in-memory processing are used to improve latency and throughput [60, 68, 114, 130]. In blockchain systems, reducing network latency and enhancing transaction speed remain central to performance optimisation [115].

**Data Centre Network Metrics** encompass data-centre–specific indicators, such as *Diameter Stretch*, *Path Stretch*, *Maximum Relative Size*, and *Network Utilisation* [119]. These are key in evaluating network efficiency, redundancy, and fault tolerance within large-scale data infrastructures.

## 5.3 Energy Efficiency & Optimisation

This cluster encompasses metrics and KPIs that monitor energy consumption while ensuring optimal system performance. Given the breadth and specificity of metrics in this domain, we dedicate only a high-level overview here, leaving detailed analysis for a separate paper. A description and references are provided in Appendix B, with dependencies defined through expert judgment and information drawn from [78, 110, 119, 129, 152, 168]. Some metrics were omitted or merged with others due to limited coverage (e.g., *H-POM*, *SI-POM*, *OSWE*, *PESavings*), lack of operational relevance (e.g., *Technology Carbon Efficiency (TCE)*), or broader applicability outside of data centres (e.g., *Material Recycling Ratio (MRR)*).

The most widely adopted metric for data centres is *Power Usage Effectiveness (PUE)*, defined as the ratio of total facility energy use to the energy consumed by computing equipment. PUE is standardised in ISO/IEC 30134-2 and EN 50600-4-2, and remains a cornerstone for benchmarking data centre efficiency. For further details on energy efficiency metrics, we refer readers to the comprehensive survey by Reddy et al. [119], as well as the ISO/IEC 30134 and EN 50600-4-x standards, which cover the majority of metrics identified in the literature.

In summary, metrics within the *Operational Efficiency* cluster play a critical role in linking system performance to data valuation and monetisation strategies. By ensuring resilient, efficient, and optimised operations, organisations can reduce operational costs, increase data reliability, and improve service continuity—factors that directly enhance the intrinsic and perceived value of data assets. Furthermore, operational efficiency supports sustainable data management practices by aligning performance optimisation with energy efficiency and risk mitigation, ultimately reinforcing both the financial and strategic dimensions of data value creation.

## 6 DISCUSSION

In this section, we present our reflections on the domain of data valuation and data monetisation, outlining the limitations of our study and highlighting open challenges that remain unaddressed in the literature.

As discussed throughout this work, we have conducted an extensive survey and sought to include as many metrics and KPIs as possible within the context of data valuation and data monetisation. Our long-term objective is to move toward a standardised framework capable of encompassing every aspect of an organisation's business operations. We acknowledge, however, the complexity of establishing such a universal framework and recognise the substantial efforts already undertaken by the research community in developing comprehensive data quality frameworks [98, 128]. By narrowing our focus to valuation and monetisation, it is possible that some metrics applicable to these domains were inadvertently omitted.

A major challenge in developing a unified framework stems from the significant overlap among metrics, many of which exhibit similar definitions or operational purposes (e.g., *Timeliness* vs. *Currency*). In constructing our taxonomy, we merged metrics that were conceptually equivalent or inversely related and could thus be derived from one another. For example, we grouped *Clarity* and *Understandability* under a single concept, as both address the prevention of ambiguity in data use. While such aggregation introduces an element of subjectivity, it enhances consistency and interpretability within the taxonomy. Similar approaches have been adopted by [98] and [128], who also consolidated synonymous or overlapping metrics within their frameworks. Nonetheless, these definitional discrepancies continue to impede the establishment of a universally accepted standard, contributing to the proliferation of heterogeneous data quality frameworks over the years [97]. This fragmentation presents a significant barrier to the development of a truly global data market—one in which organisations can reliably and equitably assess both the value of their own data and that of external data assets.

Another ongoing challenge arises from the continuous emergence of new metrics, often driven by advances in specialised domains. For instance, recent developments in artificial intelligence have led to the introduction of metrics such as *Data Purity* [15], *Class Parity*, *Data Fairness*, and *Label Noise*. While these metrics address legitimate needs, they also complicate the maintenance of a stable and comprehensive framework. A standardised taxonomy must therefore remain adaptive—capable of incorporating and harmonising newly introduced metrics while preserving conceptual coherence across industries. Achieving this goal will require ongoing monitoring, normalisation, and standardisation efforts by both academia and industry.

A further challenge lies in identifying and formalising the relationships among the many metrics uncovered in this study. While some metrics have explicit dependencies (e.g., *Availability* relies on *Integrity* and *Accessibility*), others are linked more implicitly through shared operational or strategic objectives. These interconnections are rarely fixed: they can vary significantly depending on the application domain, organisational context, and decision-making framework. For instance, *Accuracy* may directly influence value creation in a financial context, whereas in IoT environments, *Timeliness* or *Reliability* may take precedence. Understanding and mapping these dynamic relationships is therefore essential to operationalise the taxonomy and to support context-aware valuation methodologies.

Finally, it is important to emphasise that although the taxonomy presented here encompasses hundreds of metrics and KPIs, most organisations will apply only a subset of them. The taxonomy is not intended to be prescriptive or exhaustive in a practical sense, but rather to delineate the full realm of possibilities for those seeking to understand, measure, and monetise their data. Its role is to provide a comprehensive reference framework from which stakeholders can select relevant metrics aligned with their strategic objectives, data maturity, and industry context.

With these challenges in mind, our future work will focus on refining the taxonomy presented here and elucidating the relationships among its metrics. In particular, we plan to map interdependencies between metrics and demonstrate how they are currently employed across valuation and decision-making contexts. We will also illustrate how this taxonomy is implemented within a decision-support framework to assist stakeholders in identifying the most relevant metrics based on strategic objectives, data maturity, and organisational priorities. Ultimately, we aim to facilitate the convergence toward a dynamic and adaptive standard for data valuation and monetisation that evolves alongside technological and organisational innovation.

## 7 CONCLUSION

In this study, we conducted an extensive systematic literature review of key performance indicators (KPIs) and metrics relevant to data valuation and data monetisation, focusing in particular on the domains of *Data Quality*, *Governance & Compliance*, and *Operational Efficiency*. This work forms part of a broader research effort [156] examining every dimension of organisational performance through the lens of the Balanced Scorecard (BSC). Following the four BSC perspectives, we categorised hundreds of metrics and KPIs identified in the literature into a comprehensive taxonomy. In doing so, we grouped metrics into higher-level clusters that can function as composite KPIs (e.g., data quality) and identified numerous interconnections among these metrics and indicators.

Taken together, the three core clusters explored in this work—*Data Quality*, *Governance & Compliance*, and *Operational Efficiency*—form the foundational pillars of data valuation and monetisation. Each cluster captures a distinct yet interdependent dimension of how organisations create, protect, and exploit value from data. *Data Quality* ensures that information is accurate and trustworthy; *Governance & Compliance* provides the ethical and regulatory framework necessary for responsible data use; and *Operational Efficiency* translates these principles into measurable performance and sustainable value creation. By uniting these domains under a shared taxonomy, this work offers a holistic view of the data value ecosystem and establishes a foundation for the development of integrated, evidence-based valuation strategies that can evolve alongside technological and organisational innovation.

Future work will extend this taxonomy to the remaining BSC perspectives and further explore its practical applications. In particular, we will present its integration within a decision-support framework and a data valuation model currently under development as part of the European project *DATAMITE*[1], a large-scale international collaboration aimed at operationalising data value creation and monetisation across sectors.

## ACKNOWLEDGMENTS

This research was partially supported by the EU's Horizon Digital, Industry, and Spac program under grant agreement ID 101092989-DATAMITE. Additionally, we acknowledge Science Foundation Ireland under Grant No. 12/RC/2289 for funding the Insight Centre of Data Analytics (which is co-funded under the European Regional Development Fund).

---

[1] https://datamite-horizon.eu/